\documentclass[a4paper,12pt]{article}
\usepackage{amssymb,amsmath}

\newcommand{\RR}{{\mathbb{R}}}
\newcommand{\CC}{{\mathbb{C}}}

\newcommand{\pa}{\partial}
\newcommand{\ii}{{\rm i}}
\newcommand{\dd}{{\rm d}}

\newcommand{\sfrac}[2]{{\textstyle\frac{#1}{#2}}}
\newcommand{\Tr}{\mathrm{Tr}}

\newcommand{\g}{\mathfrak{g}}
\newcommand{\h}{\mathfrak{h}}
\newcommand{\m}{\mathfrak{m}}
\newcommand{\kk}{\mathfrak{k}}

\title{The Skyrme model and \\ chiral perturbation theory}
\author{Derek Harland\footnote{email address: d.g.harland@leeds.ac.uk}
  \bigskip
  \\School of Mathematics,
  \\University of Leeds,
  \\LS2 9JT}
\date{21st December 2016}

\begin{document}

\maketitle

\begin{abstract}
A lagrangian which describes interactions between a soliton and a background field is derived for sigma models whose target is a symmetric space.  The background field modifies the usual moduli space approximation to soliton dynamics in two ways: by introducing a potential energy, and by inducing a Kaluza-Klein metric on the moduli space.  In the particular case of the Skyrme model, this lagrangian is quantised and shown to agree with the leading pion-nucleon term in the chiral effective lagrangian, which is widely used in theoretical nuclear physics.  Thus chiral perturbation theory could be considered a low energy limit of the Skyrme model.
\end{abstract}

\section{Introduction}

The Skyrme model and chiral perturbation theory are two alternative models of nuclear physics (for reviews, see \cite{skyrmereview,cptreview}).  They display some superficial similarities: both model pions using a nonlinear sigma model with target SU(2), and chiral symmetry plays an important role in both.  However, baryons are treated very differently in the two models: in chiral perturbation theory baryons are quantum excitations of one of the fields in the lagrangian, whereas in the Skyrme model baryons emerge naturally as topological excitations of the pion field, known as skyrmions.

The purpose of this article is to understand how the theories are related.  Our claim is that the chiral lagrangian is an effective description of the Skyrme model, valid at low energies and large separations.  Demonstrating this claim has involved the development of a sophisticated understanding of the interactions of Skyrme fields, about which more will be said momentarily.  Our results may be viewed as a counterpart to Coleman's duality between the quantised sine-Gordon and massless Thirring models \cite{coleman75}: in both our and Coleman's work the quantised dynamics of solitons are related to the dynamics of a fermionic field.  However, there are some important differences: our result is based on approximations and effective field theories, whereas Coleman's is an exact duality; also, our result applies in a wide range of field theories, whereas Coleman's is rather specialised.

As the contents of the paper are rather technical we will try to summarise them here.  Our strategy for comparing the Skyrme model and the chiral effective lagrangian is to derive from both a Schr\"odinger equation for a nucleon, and to show that the two equations agree.  Our method for deriving a Schr\"odinger equation from the chiral effective lagrangian in section 5 is rather elementary: from the leading pion-nucleon term in the lagrangian we write down a Dirac equation for a nucleon, and from this derive a Schr\"odinger equation in the same way that one derives a Schr\"odinger equation for an electron.  The reason for using only the leading term in the chiral effective lagrangian is that this is the level of accuracy to which we are able to work in the Skyrme model; it would certainly be interesting to compare subleading terms, but that is beyond the scope of this paper.

The calculations from the Skyrme model are rather more substantial.  The first stage is to derive an action \eqref{worldline action} describing the classical mechanics of a skyrmion interacting with a background pion field.  This action only describes motion associated with the lightest modes of the skyrmion; heavier modes are neglected.  The interactions of skymions have been studied in various places in the literature \cite{schroers,gp94}.  Despite this, the action that we derive illuminates some features that have until now gone unnoticed: specifically, a background Skyrme field induces a Kaluza-Klein metric on the moduli space of a skyrmion.  This metric correction is needed in our calculation to produce a Schr\"odinger equation consistent with chiral symmetry; we expect it to find applications elsewhere, for example in the study of skyrmion scattering.

Our method for deriving the action \eqref{worldline action} involves several new ideas.  We develop in section 2 a general formalism for sigma models whose target is a symmetric space.  This has two advantages: first, it makes our derivation valid for a large class of field theories, including not only the Skyrme model but also the baby Skyrme and Faddeev models (interactions of solitons in the latter two have been studied in \cite{psz95,ward00}).  Second, it makes the appearance of a gauge field and Kaluza-Klein metric more transparent; we suspect that the reason the Kaluza-Klein metric has gone unnoticed until now is a lack of transparency in the usual formulation of the Skyrme model.  Another new idea is our ansatz, introduced in section 3, for superposing Skyrme fields; while it superficially appears similar to the so-called ``product ansatz'' it is not the same.  This ansatz is rather natural in the symmetric space formulation, and seems to give a cleaner description for the dynamics of interaction Skyrme fields.

Many studies of Skyrme interactions use tools such as the product ansatz without justification.  A distinction of our calculation is that we are able to justify it, at least heuristically.  By treating one of the two Skyrme fields as a small ``background'', we are able to motivate our calculation of the action \eqref{worldline action} along similar lines to Manton's moduli space approximation \cite{manton}.

From the classical action \eqref{worldline action} we derive in section 4 a Schr\"odinger equation for a wavefunction on the skyrmion moduli space.  In the case of the Skyrme model this is shown using a mode expansion to reduce at low energies to a Schr\"odinger equation for a spinor-valued field.  This Schr\"odinger equation agrees with the one derived from the chiral effective lagrangian in section 5.  The methods of section 4 can be thought of as an application of ideas introduced in \cite{anw} to more general dynamics in the soliton moduli space.  We make some concluding remarks in section 6.

\section{Sigma models}

\subsection{Symmetric spaces}

This article deals with sigma models whose targets are symmetric spaces.  Therefore we begin by recalling the definitions and some properties of symmetric spaces.

Let $G$ be a compact connected Lie group and let  $\sigma:G\to G$ be a map satisfying $\sigma(gg')=\sigma(g)\sigma(g')$ and $\sigma^2(g)=g$ for all $g,g'\in G$.  Let $H\subset G$ be the fixed set of $\sigma$.  Then $H$ must be a subgroup of $G$.  The quotient space $G/H$ is called a symmetric space.  Points in $G/H$ may be represented by elements $g\in G$, and the elements $g$ and $gh$ are understood to represent the same point for any $h\in H$.

For an example, let $G=\mathrm{SU}(2)\times\mathrm{SU}(2)$ and let $\sigma(\xi_L,\xi_R)=(\xi_R,\xi_L)$ for all $\xi_L,\xi_R\in\mathrm{SU}(2)$.  Then the fixed set of $\sigma$ is $H=\{(h,h)\::\:\xi\in\mathrm{SU}(2)\}$ and this is obviously isomorphic to $\mathrm{SU}(2)$.  In this example the quotient space $G/H$ is diffeomorphic to $\mathrm{SU}(2)$; a diffeomorphism from $G/H$ to $\mathrm{SU}(2)$ is given by
\[ (\xi_L,\xi_R)\mapsto U=\xi_L\xi_R^{-1}. \]
Note that $U$ is well-defined, because $(\xi_Lh)(\xi_Rh)^{-1}=\xi_L\xi_R^{-1}$.  Similarly, any compact connected Lie group $H$ can be identified with a symmetric space $H\times H/H$.  Other examples of symmetric spaces include spheres and complex projective spaces of any dimension.

The automorphism $\sigma$ of $G$ defines a linear map $\sigma:\mathfrak{g}\to\mathfrak{g}$ of the Lie algebra $\mathfrak{g}$, which is also denoted $\sigma$.  There are natural relationships between the two maps:
\begin{align*}
[\sigma(X),\sigma(Y)] &= \sigma([X,Y])&\forall X,Y\in\g,\\
\sigma(gXg^{-1}) &= \sigma(g)\sigma(X)\sigma(g^{-1})&\forall X\in\g,\, g\in G,\\
\sigma\big(g^{-1}(t)\pa_t g(t)\big) &= \sigma\big(g^{-1}(t)\big)\pa_t \sigma\big(g(t)\big)& \forall g:\RR\to G.
\end{align*}
The map $\sigma:\g\to\g$ is linear and squares to 1, so its eigenvalues can be either 1 or -1.  The corresponding eigenspaces are denoted $\mathfrak{h}$ and $\mathfrak{m}$.  The space $\mathfrak{h}$ is the Lie algebra of the subgroup $H$, and the space $\m$ can be identified with the tangent space of $G/H$ at the point represented by the identity element $1_G$.  There is a direct sum decomposition $\mathfrak{g}=\mathfrak{h}\oplus\mathfrak{m}$ and hence projection maps $\pi_\m:\g\to\m$, $\pi_\h:\g\to\h$.  Since $\sigma$ respects the Lie bracket it holds that
\[ [\h,\h]\subset\h,\quad [\h,\m]\subset\m,\quad[\m,\m]\subset \h. \]
In what follows $I_\alpha$ will denote a basis for $\h$ and $I_a$ a basis for $\m$.  The structure constants will be denoted $f_{\ast\ast}^\ast$ and are defined by $[I_\alpha,I_a]=f_{\alpha a}^bI_b$ etc.  Note that $f_{\alpha a}^\beta=0$ due to the above relations.

In our example with $G=\mathrm{SU}(2)\times\mathrm{SU}(2)$ and $H\cong\mathrm{SU}(2)$, the action of $\sigma$ on the Lie algebra is $\sigma(X_L,X_R)=\sigma(X_R,X_L)$ for all $X_L,X_R\in\mathfrak{su}(2)$.  A basis for $\h$ is given by
\[ I_\alpha = \left( -\frac\ii2\sigma_\alpha,-\frac\ii2\sigma_\alpha\right),\quad \alpha=1,2,3, \]
where $\sigma_a$ are the Pauli matrices.  A basis for $\m$ is given by
\[ I_a = \left( \frac\ii2\sigma_a,-\frac\ii2\sigma_a\right),\quad a=1,2,3. \]

\subsection{Lagrangians}

The degrees of freedom in our field theories will be maps $\phi:\RR^{d,1}\to G/H$.  These can be represented by maps $g:\RR^{d,1}\to G$.  Two maps $g(x),\,g'(x)=g(x)h(x)$ related by a gauge transformation $h:\RR^{d,1}\to H$ represent the same map $\phi$ so are understood to be equivalent.  In what follows we will make little distinction between the map $\phi$ and the $G$-valued map $g$ that represents it.

Given such a map $g$, we write
\[ L_\mu = \pi_\m(g^{-1}\pa_\mu g),\quad A_\mu = \pi_\h(g^{-1}\dd g). \]
Under gauge transformations these transform as
\begin{equation}
\label{gauge transformation}
g(x)\mapsto g(x)h(x),\quad L_\mu \mapsto h^{-1}L_\mu h,\quad A_\mu \mapsto h^{-1}\pa_\mu h + h^{-1}A_\mu h
\end{equation}
because the adjoint action of $H$ on $\g$ respects the splitting $\g=\h\oplus\m$.  Thus $A_\mu$ is a gauge field and $L_\mu$ is a differential one-form; $A_\mu$ can in fact be identified with the pull-back to spacetime of the Levi-Civita connection on the tangent bundle of $G/H$, and $L_\mu$ represents the differential $\dd\phi$ of the map $\phi$.

We consider field theories defined by actions of the form
\begin{equation}\label{action}
S[g] = \int\big\{\mathcal{L}(L_\mu) - \mathcal{U}(g)\big\}\dd^{d+1}x.
\end{equation}
We suppose that function $\mathcal{L}(L_\mu)$ is invariant under gauge transformations and Lorentz transformations, and also satisfies $\mathcal{L}(-L_0,L_i)=\mathcal{L}(L_0,L_i)$ and $\mathcal{L}(L_0,-L_i)=\mathcal{L}(L_0,L_i)$.  We assume that the potential function $\mathcal{U}:G\to\RR$ is invariant under gauge transformations, $\mathcal{U}(gh)=\mathcal{U}(g)$, and also satisfies $\mathcal{U}(hg)=\mathcal{U}(g)$ for all $g\in G$, $h\in H$.  It follows that the action \eqref{action} is invariant under gauge transformations \eqref{gauge transformation}, Lorentz transformations, time reversal $g(x^0,\mathbf{x})\mapsto g(-x^0,\mathbf{x})$, space reversal $g(x^0,\mathbf{x})\mapsto g(x^0,-\mathbf{x})$, the discrete transformation $g\mapsto \sigma(g)$, and transformations of the form
\[ g(x)\mapsto h g(x),\quad h\in H. \]
Borrowing terminology from the Skyrme model, the latter will be referred to as ``isorotations'' and the group $H$ as the ``isospin group''.  Note that isorotations differ from gauge transformations in that $H$ acts on $g$ from the left rather than the right.  If the potential function $\mathcal{U}$ is absent then the action enjoys additional invariance under $g(x)\mapsto g' g(x)$ for any $g'\in G$.

We assume that $\mathcal{U}$ attains a minimal value 0 at the point $\phi_0\in G/H$ represented by the identity element $1_G\in G$, and refer to this point as the vacuum.  For fields $g=\exp(Y)$ close to the vacuum, we assume that the lagrangian takes the form
\begin{equation}
\label{linearised lagrangian}
\mathcal{L}(L_\mu)-\mathcal{U}(g) = -\frac12\kappa_{ab}( \pa_\mu Y^a\pa^\mu Y^b + m^2 Y^aY^b) + O(Y^4),
\end{equation}
with $\kappa_{ab}$ a symmetric tensor that defines a non-degenerate $H$-invariant metric $\langle Y,Y\rangle=\kappa_{ab}Y^a Y^b$ on $\m$ and $m\geq 0$.

For later use we record here the equations of motion derived from the lagrangian.  Under a small variation $g\mapsto g\exp(Y)$ the derived field $L_\mu$ transforms as
\[ L_\mu \mapsto L_\mu + \pi_\m(\pa_\mu Y + [A_\mu,Y]) + O(Y^2). \]
Therefore, writing $L_\mu = L_\mu ^aI_a$ and $A_\mu = A_\mu^\alpha I_\alpha$, the Euler-Lagrange equation derived from the action is
\[ \frac{\pa}{\pa x^\mu}\frac{\pa \mathcal{L}}{\pa L_\mu^a} - f_{\alpha a}^b A_\mu^\alpha\frac{\pa \mathcal{L}}{\pa L_\mu^b} + \delta_a \mathcal{U} = 0, \]
where $\delta_a\mathcal{U}(g):=\dd\mathcal{U}(g \exp(tI_a))/\dd t|_{t=0}$.

The Skyrme model is one field theory that can be described within our framework.  Here $G=\mathrm{SU}(2)\times\mathrm{SU}(2)$, $H$ is the diagonal subgroup isomorphic to $\mathrm{SU}(2)$, and $g(x)=(\xi_L(x),\xi_R(x))$.  In the Skyrme literature it is usual to work with the Skyrme field $U=\xi_L\xi_R^{-1}$, which is invariant under gauge transformations \eqref{gauge transformation}.  One has
\begin{align*}
A_\mu &= (a_\mu,a_\mu) = A_\mu^\alpha I_\alpha, \\
a_\mu &= \frac{1}{2}\big(\xi_L^{-1}\pa_\mu\xi_L+\xi_R^{-1}\pa_\mu\xi_R\big) = A_\mu^\alpha \frac{\sigma_\alpha}{2\ii}, \\
L_\mu &= \frac{1}{2}(\ell_\mu,-\ell_\mu)\in\mathfrak{m} = L_\mu^a I_a, \\
\ell_\mu &= \xi_L^{-1}\pa_\mu\xi_L-\xi_R^{-1}\pa_\mu\xi_R = L_\mu^a \ii\sigma_a,
\end{align*}
such that $g^{-1}\pa_\mu g=A_\mu+L_\mu$.  In the gauge where $\xi_R=1$ these expressions reduce to $2a_\mu=\ell_\mu= U^{-1}\pa_\mu U$, while in the gauge where $\xi_L=1$ they reduce to $-2a_\mu=\ell_\mu= \pa_\mu UU^{-1}$.  Most of the Skyrme literature implicitly adopts one or the other of these gauges, both of which suffer the disadvantage of $a_\mu$ and $\ell_\mu$ being difficult to tell apart.  The standard Skyrme lagrangian takes the form
\begin{equation*}
\mathcal{L} = \frac{F_\pi^2}{16\hbar}\Tr(\ell_\mu\ell^\mu) + \frac{\hbar}{32e^2}\Tr([\ell_\mu,\ell_\nu][\ell^\mu,\ell^\nu]) 
- \frac{F_\pi^2m_\pi^2}{8\hbar^3}\Tr(1-U).
\end{equation*}
Here $F_\pi$ is the pion decay constant and $m_\pi$ is the pion mass, whose measured values are approximately 184MeV and 137MeV, and $e$ is a dimensionless parameter.  We have explicitly included $\hbar$ as it is convenient to use units where $\hbar\neq1$.  The vacuum is clearly $U=1$ and for $U$ close to the vacuum the lagrangian takes the form described in \eqref{linearised lagrangian}, with 
\[ \kappa_{ab} = \frac{F_\pi^2}{8\hbar}\delta_{ab},\quad m=\frac{m_\pi}{\hbar} \]
with respect to the basis for $\m$ introduced above.  Other field theories that fit our framework include the baby Skyrme model and the Faddeev model.

\subsection{Solitons}

We will assume that the theory described by the action \eqref{action} supports a soliton, i.e.\ a static, stable, spatially-localised, non-trivial solution of its equations of motion.  The Skyrme model, baby Skyrme model, and Faddeev model are all examples of theories that support solitons.  In all of these theories solitons carry a topological charge, but this will have little bearing on our discussion.

The action \eqref{action} has a conserved energy; when evaluated on a static field $g:\RR^d\to G$ this energy takes the form,
\[ E[g]:= \int_{\RR^d}\big\{ \mathcal{V}(L_i) + \mathcal{U}(g)\big\} \dd^d\mathbf{x}, \]
where we have written $\mathcal{V}(L_i)=-\mathcal{L}(L_0=0,L_i)$.  A soliton is a map $g_S$ which is a local minimum of $E[g]$.  We assume that solitons are spatially-localised in the sense that for large $\mathbf{x}$
\begin{equation}
\label{soliton asymptotics}
g_S(\mathbf{x})\sim \exp(Q^i\pa_iG(\mathbf{x}))
\end{equation}
up to gauge equivalence, where $Q^i\in\m$ for $i=1,\ldots,d$ and $G(\mathbf{x})$ is the Greens function satisfying
\[ \triangle G(\mathbf{x})-m^2 G(\mathbf{x}) = \delta(\mathbf{x}). \]
For example, when $d=3$, $G(\mathbf{x})=-e^{-m|\mathbf{x}|}/4\pi|\mathbf{x}|$.  We also assume that the soliton has the discrete symmetry
\begin{equation}
\label{inversion symmetry}
\sigma(g_S(-\mathbf{x}))=g_S(\mathbf{x}).
\end{equation}
Note that this symmetry is compatible with the asymptotics \eqref{soliton asymptotics}.

The \emph{moduli space} of a soliton is the set of fields $g(\mathbf{x})$ which are also local minima of $E[g]$ and which are degenerate in energy with the soliton.  Applying a rotation, isorotation, or translation to a soliton does not change its energy, so the images of $g_S$ under these transformations form at least part of the moduli space.  We will in fact assume that they constitute the whole of the moduli space, as is the case in the Skyrme, baby Skyrme and Faddeev models.  Thus the moduli space consists of fields of the form
\[ hg_S(R^{-1}(\mathbf{x}-\mathbf{X})). \]
where $(h,R,\mathbf{X})\in H\times \mathrm{SO}(d)\times \RR^d$.  As a manifold, the moduli space is diffeomorphic to
\[ (H\times \mathrm{SO}(d))/K \times \RR^d, \]
where $K\subset H\times\mathrm{SO}(d)$ is the subgroup consisting of pairs $(h,R)$ such that $hg_S(R^{-1}\mathbf{x})=g_S(\mathbf{x})$.

Two important quantities associated with a soliton are its \emph{rest mass} and its \emph{moment of inertia}.  The rest mass $M_S$ is simply the classical energy of the soliton:
\[ M_S:= \int_{\RR^d} \big\{ \mathcal{V}(L_i^S) + \mathcal{U}(g_S)\big\} \dd^d\mathbf{x}, \]
in which $L_i^S=g_S^{-1}\pa_i g_S$.  The moment of inertia describes the response of the soliton to rotation and isorotation.  Consider a rigidly isorotating rotating soliton of the form
\[ g(x^0,\mathbf{x})=\exp(Zt)g_S(\exp(-Wt)\mathbf{x}), \]
with $Z\in\h$ and $W\in\mathfrak{so}(d)$.  The moment of inertia of the soliton is the quadratic form $\Lambda_S$ on $\h\oplus\mathfrak{so}(d)$ defined by
\[
\frac12\Lambda_S(Z,W) +O((Z,W)^3) = \int_{\RR^d} \mathcal{T}(L_0,L_i)|_{x^0=0}\dd^d \mathbf{x},
\]
where $\mathcal{T}(L_0,L_i):=\mathcal{L}(L_0,L_i)-\mathcal{V}(L_i)$.  The physical interpretation is that $\frac12\Lambda_S(Z,W)$ is the leading contribution to the kinetic energy of the  soliton.  By inserting the expressions
\[ L_i = \pi_\mathfrak{m}(g_S^{-1}\pa_i g_S)=L_i^S,\quad L_0 = \pi_\m(g_S^{-1}Zg_S) - W^i_jx^jL_i^S, \]
one obtains the following formula for $\Lambda_S$:
\begin{equation}
\label{inertia tensor}
\frac12\Lambda_S(Z,W) = \int_{\RR^d} \mathcal{T}^{(2)}(\pi_\m(g_S^{-1}Zg_S) - W^i_jx^jL_i^S,L_i^S)\dd^d \mathbf{x},
\end{equation}
where $\mathcal{T}^{(2)}$ denotes the part of $\mathcal{T}$ which is quadratic in $L_0$.  Note that $\Lambda_S$ is invariant under the adjoint action of $K$ on $\h\oplus\mathfrak{so}(d)$, because both $g_S$ and $\mathcal{T}$ are $K$-invariant.  Note also that $\Lambda_S$ vanishes on the Lie algebra $\mathfrak{k}$ of $K$ and is non-negative on $(\h\oplus\mathfrak{so}(d)/\mathfrak{k}$.
It is well-known in the theory of homogeneous spaces that any such $\Lambda_S$ defines a metric on $(H\times \mathrm{SO}(d))/K$.

In the case of the Skyrme model the soliton is known as a skyrmion and is of hedgehog form:
\[ U_S(\mathbf{x}) = \exp\left(\ii f\left(\frac{eF_\pi|\mathbf{x}|}{2\hbar}\right) \frac{x^i\sigma_i}{|\mathbf{x}|}\right). \]
Note that for convenience we have arranged for the argument $r=eF_\pi|\mathbf{x}|/2\hbar$ of $f$ to be dimensionless.  The field $U_S$ is invariant under simultaneous rotation and isorotation, so $K$ is a diagonal subgroup of $\mathrm{SU}(2)\times\mathrm{SO}(3)$ which is isomorphic to $\mathrm{SU}(2)$.  The moduli space is therefore diffeomorphic to $\mathrm{SO}(3)\times \RR^3$.

The mass of the skyrmion is
\begin{align*}
 M_S &= 4\pi \frac{F_\pi}{4e} \mu[f] \\
 \mu[f] &= \int_0^\infty \left\{(f')^2 + 2\frac{\sin^2f}{r^2} + 2\frac{(f')^2\sin^2f}{r^2} + \frac{\sin^4f}{r^4} + 2\bar{m}^2(1-\cos f)\right\}r^2\dd r,
\end{align*}
where $\bar{m}=2m_\pi/F_\pi e$.
The function $f$ should be chosen to minimise $\mu[f]$, subject to the boundary conditions that $f(0)=\pi$ and $f(r)\sim q(1/r^2+m/r)\exp(-mr)$ as $r\to\infty$ for some $q>0$.  The dipole coefficient is therefore
\[ Q^i = 4\pi q\left(\frac{2\hbar}{F_\pi e}\right)^2\delta^{ia}I_a. \]
Writing $Z=Z^\alpha I_\alpha$ and $W=W^iJ_i$, with $J_i\in\mathfrak{so}(3)$ the matrix $(J_i)_{jk}=-\epsilon_{ijk}$, the moment of inertia tensor is
\begin{align}
\label{skyrmion inertia tensor}
 \Lambda(Z,W) &= \frac{16\pi}{3}\frac{\hbar^2}{F_\pi e^3}\lambda[f] (Z^i-W^i)(Z^i-W^i) \\
\nonumber
 \lambda[f] &= \int_0^\infty \left\{1+(f')^2 + \frac{\sin^2f}{r^2}\right\}\sin^2f\,r^2\dd r.
\end{align}
Note that $\Lambda(Z,W)=0$ if $Z^i=W^i$, reflecting the isospin-spin symmetry of the soliton.

\section{Soliton interacting with a background}

In this section we will derive an effective action for a soliton moving in a background field $\bar{\phi}$.  We have in mind that the background is induced by sources or boundary conditions, and does not contain any solitons itself.  Our approach to deriving this effective action is based on a modest modification of Manton's moduli space approximation \cite{manton}, so we begin by recalling the main ideas behind that.

\subsection{Moduli space approximation with potential}

The moduli space approximation is a way to approximate the dynamics of the lightest modes of a soliton, namely the zero-modes.  It can be explained by making an analogy with the mechanics of a particle moving in a potential $V(\mathbf{x})$.  The moduli space is the set of vacua of $V$ (that is, the set of values of $\mathbf{x}$ that minimise $V$), and we suppose that moduli space forms a curve or surface.  If for example $V_0(x_1,x_2)=(1-x_1^2-x_2^2)^2$ then the moduli space is a unit circle.  The moduli space approximation says that a particle which is initially in the moduli space and whose initial velocity is tangent to the moduli space will remain in the moduli space.  This is a good approximation to the dynamics if the initial velocity is small, because in that situation there is not enough energy in the system for the soliton to stray far from the moduli space.  The moduli space approximation is in effect an adiabatic approximation: it divides the degrees of freedom of the particle into those tangential and normal to the moduli space, and says that the normal degrees of freedom are not excited as the tangential degrees of freedom vary slowly.

If the potential $V$ is perturbed slightly then the moduli space approximation should still be reliable.  For example, if the potential given above is replaced by $V_\epsilon(x_1,x_2) = (1-x_1^2-x_2^2)^2+\epsilon x_1$, with $\epsilon\ll 1$, then under the same assumptions as before a particle is not expected to stray far from the circle.  This is the modification of the moduli space approximation that we will use to described a soliton interacting with a background.

The moduli space approximation and the modification described above provide a convenient way to derive an action principle for low-energy motion.  We work within the framework of lagrangian mechanics; thus the path $\mathbf{X}(t)$ followed by the particle must be a criticial point of an action $S$, which is a function on the space $P$ of all paths in Euclidean space.  Let $P'\subset P$ denote the set of paths which remain in the moduli space and let $S'$ denote the restriction of $S$ to $P'$.  For a path to be a critical point of $S'$ it must be stable to variations of the path tangent to the moduli space, whereas to be a critical point of $S$ it must be stable to both tangential and normal variations.  The moduli space approximation says that paths which are stable to tangential variations are also to a good approximation stable to normal variations.  Therefore critical points of $S'$ are good approximations to critical points of $S$.  Thus the action $S'$ obtained by evaluating $S$ on paths in the moduli space is an effective action for low-energy dynamics.

\subsection{Worldline action}

We now consider a soliton moving in a background.  We assume from now on that the background field $\bar{\phi}$ is induced by a source $J^\mu_a(x)$ localised in space and time.  Then $\bar{\phi}$ (or rather, its representative $\bar{g}$) solves the equations of motion derived from the action
\[ S_J[g]=S[g]+\int J^\mu_a L_\mu^a \dd^{d+1}x.\]
The equations of motion solved by $\bar{g}$ then take the form
\begin{equation}
\label{background eom}
\pa_\mu \left(\frac{\pa \mathcal{L}}{\pa L_\mu^a}(\bar{L})+J^\mu_a\right) - \bar{A}_{\mu}^\alpha f_{\alpha a}^b\frac{\pa \mathcal{L}}{\pa L_\mu^b}(\bar{L}) + \delta_a \mathcal{U}(\bar{g}) = 0.
\end{equation}
Let $\bar{\phi}(x)$ to be the (presumed unique) stable solution  such that far from the source $\bar{\phi}$ is close to the vacuum $\phi_0$ and its derivatives are small.  As the source $J^\mu_a$ is slowly turned off this solution is assumed to approach the vacuum solution $\bar{\phi}(x)=\phi_0$.  We choose $\bar{g}(x)$ to be a representative of $\bar{\phi}$ such that far from the source $g\approx1_G$ and $\bar{A}_\mu=\pi_\h(\bar{g}^{-1}\pa_\mu g)$ and its derivatives are small.

We wish to treat $\bar{\phi}$ as a background and study fluctuations about it.  A convenient way to do so is to introduce
\[\tilde{g}:=\bar{g}^{-1} g.\]
Then the background $g=\bar{g}$ corresponds to $\tilde{g}=1$.  Since $\tilde{g}(x)\mapsto\tilde{g}(x)h(x)$ under gauge transformations $g(x)\mapsto g(x)h(x)$, $\tilde{g}$ represents a well-defined map $\tilde{\phi}:\RR^{d,1}\to G/H$.  The action that describes interactions of $\tilde{\phi}$ with the background field is
\[ S_B[\tilde{g}]:= S_J[\bar{g}\tilde{g}] - S_J[\bar{g}]. \]
When $\bar{g}=1$ the action $S_B$ equals the original action $S$, so $S_B$ is a perturbation of the original action, just as $V_\epsilon$ was a perturbation of the potential $V_0$ in our example above.

In the calculation that follows, the field $\tilde{g}$ will contain a soliton and the field $\bar{g}$ will be the background field with which it interacts.  Quantities associated with $\tilde{g}$ will be decorated with a $\tilde{}$ and those associated with $\bar{g}$ with a $\bar{}$ (for example $\bar{L}_\mu=\pi_\m(\bar{g}^{-1}\pa_\mu\bar{g})$).  The combination $g=\bar{g}\tilde{g}$ on which the action $S_J$ is evaluated resembles the ``product ansatz'' employed in \cite{schroers,gp94} and elsewhere, but is not the same: in Skyrme language the formula $g=\bar{g}\tilde{g}$ is equivalent to $U = \bar{\xi}_L\tilde{U}\bar{\xi}_R$, where $\bar{\xi}_L\bar{\xi}_R=\bar{U}$, whereas the product ansatz is $U=\bar{U}\tilde{U}$ or $\tilde{U}\bar{U}$.

An interesting feature of the action $S_B[\tilde{g}]$ defined above is that it agrees with $S$ to leading order, so is in this sense a small perturbation of $S$.  To see this, suppose that both $\tilde g$ and $\bar{g}$ are close to their vacua, i.e.\ that $\bar{g}=\exp(\bar{Y})$ and $\tilde{g}=\exp(\tilde{Y})$, with $\bar{Y},\tilde{Y}$ and their derivatives small.  We consider a series expansion in $\tilde{Y}$.  If $\tilde{Y}=0$ then $S_B[\tilde{g}]=0$, so the constant term in the expansion is zero.  The linear term in the expansion vanishes because $\bar{g}$ satisfies the equations of motion for $S_J$.  Therefore the first non-zero term is quadratic in $\tilde{Y}$.  Since both $\tilde{Y}$ and $\bar{Y}$ are small the quadratic term in $\tilde{Y}$ can be evaluated using the expression \eqref{linearised lagrangian}; one finds that
\[ S_B \approx \int\left[-\frac12\kappa_{ab}\pa_\mu \tilde{Y}^a\pa^\mu \tilde{Y}^b - \frac{m^2}{2}\kappa_{ab}\tilde{Y}^a\tilde{Y}^b \right] \dd^{d+1}x \]
in agreement with \eqref{linearised lagrangian}.

Below we will describe in detail how to evaluate the action $S_B$ on paths in the soliton moduli space.   A path in the soliton moduli space may written $(h_S(t),R(t),\mathbf{X}(t))$.  The soliton worldline in Minkowski space is the image of the map $t\mapsto (t,\mathbf{X}(t))$.  It will prove convenient to parametrise the path not by inertial time $t$ but by proper time $\tau$ along the soliton worldline, so that the path is $\tau\mapsto(h_S(\tau),R(\tau),X^\mu(\tau))$ and $\pa_\tau X^\mu\pa_\tau X_\mu=-1$.  We claim that evaluating the action $S_B$ on a field $\tilde{g}$ carrying a soliton following a path in the moduli space results in
\begin{equation}
\label{worldline action}
S_B \approx \int\big[-M_S + \sfrac12\Lambda(h_S^{-1}D_\tau h_S,R^{-1}\pa_\tau R) + \kappa_{ab}m_i^\mu (h_SQ^ih_S^{-1})^a \bar{L}_\mu^b\big]\,\dd\tau,
\end{equation}
where we have introduced the notation
\[ D_\tau h_S(\tau):=\pa_\tau h_S(\tau)+\pa_\tau X^\mu(\tau) \bar{A}_\mu(X^\nu(\tau)) h_S(\tau) \]
and
\begin{equation}
\label{worldline boost}
m_0^0=\pa_\tau X^0, \, m_0^i=m_i^0=\pa_\tau X^i,\, m_i^j=\delta_i^j+\frac{\pa_\tau X^i\pa_\tau X^j}{1+\pa_\tau X^0} .
\end{equation}
The matrix $m_\mu^\nu$ is a Lorentz boost which maps $(1,0,\ldots,0)$ to $\pa_\tau X^\mu$; the columns $m_i^\mu$ form an orthonormal frame for the normal bundle of the wordline.  The action \eqref{worldline action} is invariant under Poincar\'e transformations.  It is also invariant under gauge transformations $\bar{h}(x)$ acting as
\[ \bar{g}(x)\mapsto\bar{g}(x)\bar{h}(x),\quad h_S(\tau)\mapsto  \bar{h}(X^\mu(\tau))^{-1}h_S(\tau). \]
This action \eqref{worldline action} should be a reliable description of the interaction of a slowly-moving soliton with a weak background field.

The physical interpretation of the action \eqref{worldline action} is that it describes an oriented particle with scalar dipole charges interacting with the background field $\bar{\phi}$.  The first term in the action is the standard action for a particle of mass $M_S$.  In a frame in which the velocity is zero, the second term is the kinetic energy of the soliton.  In the same frame the final term is the potential energy resulting from interaction of the dipole charges with the background field.

It is widely accepted that skyrmions interact as charged scalar dipoles, so most features of the action \eqref{worldline action} will not be surprising to experts.  One feature that does seem to be new is the appearance of a covariant derivative in the term describing kinetic energy due to isorotation.  This can be attributed to the nonlinearity of the scalar field $\bar{\phi}$ with which the particle interacts: the dipole charges of a particle with coordinates $X^\mu$ should be considered to take values in the tangent space $T_{\bar{\phi}(X^\mu)}\Sigma$ of the target manifold $\Sigma=G/H$ at $\bar{\phi}(X^\mu)$.  The dipole charges are therefore sections of a vector bundle (the pull-back of $T\Sigma$) and so must be differentiated using a covariant derivative (in this case, the covariant derivative associated with the Levi-Civita connection on $T\Sigma$).

The appearance of a covariant derivative in the action \eqref{worldline action} implies that the background field induces a Kaluza-Klein deformation of the moduli space metric.  This is most readily seen in the case of the baby Skyrme model.  Here $d=2$, $H=\mathrm{SO}(2)$, and the group $K$ is isomorphic to $\mathrm{SO}(2)$, so that the moduli space is $S^1\times\RR^2$.  The pair $(h,R)\in\mathrm{SO}(2)\times\mathrm{SO}(2)$ can be replaced by an angle $\theta\in [0,2\pi)$.  The first two terms in \eqref{worldline action} then take the form
\[ \int \big[-M_S + \frac{\lambda_S}{2}(\pa_\tau\theta+\pa_\tau X^\mu \bar{a}_\mu)^2\big]\dd\tau \]
for some positive constant $\lambda_S$, with $\bar{a}_\mu$ being the abelian background gauge field.  This agrees up to terms of order $(\pa_\tau\theta+\pa_\tau X^\mu \bar{a}_\mu)^4$ with the action of a particle of mass $M_S$ moving in a Kaluza-Klein metric
\[ \dd s^2 = \dd X_\mu\dd X^\mu + \frac{\lambda_S}{M_S}(\dd\theta+\bar{a}_\mu\dd X^\mu)^2. \]
More generally, the metric implied by \eqref{worldline action} is a non-abelian Kaluza-Klein metric on a fibre bundle over Minkowski space whose fibre is the homogeneous space $H\times\mathrm{SO}(d)/K$.  See \cite{bl87} for a review of Kaluza-Klein metrics.

\subsection{Evaluation of the action: zero acceleration}

The result \eqref{worldline action} is derived using the following assumptions:
\begin{enumerate}
\item Near the soliton wordline the background field $\bar{g}$ takes the form $\bar{g}=\exp \bar{Y}$, with $\bar{Y}$ and $\pa_\mu \bar{Y}$ small and $\pa_\mu \bar{Y}$ roughly constant. \label{assumption 1}
\item The source $J^\mu_a$ of the background field vanishes near the soliton worldine. \label{assumption 2}
\item The acceleration of the soliton is small. \label{assumption 3}
\item The speed of rotation and isorotation of the soliton is small, and the rate of change of this speed is also small. \label{assumption 4}
\end{enumerate}
We present below a detailed derivation of this result in the case that the soliton worldline has zero acceleration.  We will then sketch a derivation in the case that the acceleration is non-zero to explain why it is important to assume that the acceleration is small.  In the case that the potential function $\mathcal{U}(\bar{g})$ in the lagrangian vanishes the list of assumptions could be reduced slightly: it would not be necessary to assume that $\bar{g}$ is close to the vacuum near the soliton, but only that $\bar{g}^{-1}\pa_\mu\bar{g}$ is small and roughly constant.

A soliton following a path in the moduli space with zero acceleration and zero velocity takes the following form (up to translation):
\[ \tilde{g}\big(x^0,\mathbf{x}\big) = h_S\big(x^0\big) g_S\big(R(x^0)^{-1} \mathbf{x}\big). \]
Since a general path with zero acceleration can be obtained from such a path by Poincar\'e transformation, and the action is Poincar\'e-invariant, it suffices to evaluate $S_B$ on fields of this form.  Now we let $g=\bar{g}\tilde{g}$ and evaluate
\[ S_B[\tilde{g}]=\int_{I\times \RR^d}\big\{ \mathcal{L}(L_\mu)-\mathcal{U}(g)+J^\mu_aL_\mu^a-\mathcal{L}(\bar{L}_\mu)+\mathcal{U}(\bar{g})-J^\mu_a\bar{L}_\mu^a\big\}\dd^{d+1}x. \]
Here we evaluate the action on a time interval $I\subset\RR$ so as to obtain a finite quantity.  This choice of interval is to some extent arbitrary, but we insist it is chosen large enough that 
\begin{equation} \label{assumption 5} \pa\mathcal{L}/\pa L_0^a(\bar{L}_\mu)+J^0_a\approx 0\mbox{ on }\pa I\times\RR^d.\end{equation}

Our strategy for evaluating the action begins by choosing a spherical domain $D\subset \RR^3$ such that the soliton is contained inside $D$ and the source term has support outside $D$.  This domain should be large enough that most of the soliton's energy is contained inside $D$.  At the same time, it should be small enough that assumptions \ref{assumption 1} and \ref{assumption 2} above hold within $I\times D$.

The calculation that follows is divided into four steps.  In Step 1, we evaluate the lagrangian density over $I\times(\RR^d\setminus D)$ by treating the soliton field as a small perturbation of the background field.  This results in a boundary integral $I_1$ over $I\times \pa D$.  In Step 2 we evaluate the potential energy terms of the lagrangian density over $I\times D$, treating the background field as a perturbation of the soliton.  This results in an integral $I_3$ which equals the mass of the soliton and a boundary integral $I_4$ over $I\times\pa D$.  In Step 3 the two boundary integrals $I_1$ and $I_4$ are evaluated using asymptotic form \eqref{soliton asymptotics} of the the soliton field.  In Step 4 the remaining terms are integrated over $I\times D$, making use of the discrete symmetry \eqref{inversion symmetry} of the soliton.

\noindent\textit{Step 1.}  Outside $D$ we treat the soliton field $\tilde{g}$ as a small perturbation of the background.  Up to gauge transformation, the soliton field takes the form $\tilde{g}= \exp(\tilde{Y})$, with
\[ \tilde{Y} \approx h_SQ^ih_S^{-1}R_i^j\pa_j G. \]
Then $S_J[g]$ is approximately equal to $S_J[\bar{g}]$ plus a term which is linear in $\tilde{Y}$.  Since $\bar{g}$ solves the equations of motion \eqref{background eom} for $S_J$, this linear term can be reduced to a boundary integral:
\begin{align*}
I_1 &= \int_{I\times (\RR^d\setminus D)} \big\{ \mathcal{L}(L_\mu)-\mathcal{U}(g)+J^\mu_aL_\mu^a-\mathcal{L}(\bar{L}_\mu)+\mathcal{U}(\bar{g})-J^\mu_a\bar{L}_\mu^a\big\}\dd^{d+1}x \\
 &= \int_{I\times(\RR^d\setminus D)}\bigg\{ \bigg(\frac{\pa\mathcal{L}}{\pa L_\mu^a}(\bar{L}) + J_\mu^a\bigg)(\pa_\mu \tilde{Y}^a + \bar{A}^\alpha_\mu f^a_{\alpha b} \tilde{Y}^b) - \pa_a\mathcal{U}(\bar{g})\tilde{Y}^a \bigg\} \dd^{d+1}x \\
 &= -\int_{I\times \pa D } \frac{\pa\mathcal{L}}{\pa L_\mu^a}(\bar{L})\tilde{Y}^a \dd \Sigma_\mu.
\end{align*}
(Here we use the notation $\dd\Sigma_\mu = (\pa/\pa x^\mu)\lrcorner \dd^{d+1}x$).  Notice that there is no term involving the source $J^\mu_a$ in the final line because we have assumed that $J^\mu_a$ vanishes near the soliton.  There is no boundary term on $\pa I\times \RR^d\setminus D$ due to eq.\ \eqref{assumption 5}, and there is no boundary term on $I\times\pa\RR^d$ due to the soliton boundary condition \eqref{soliton asymptotics}.

\noindent\textit{Step 2.}  We evaluate separately the integrals of the kinetic and potential terms inside $D$.  As above, let $\mathcal{V}(L_i)=-\mathcal{L}(L_0=0,L_i)$.  The potential terms in $S_J[g]$ are
\[ I_2 = \int_{I\times D} \{ -\mathcal{V}(L_i)-\mathcal{U}(g)+\mathcal{V}(\bar{L}_i)+\mathcal{U}(\bar{g})\} \dd^{d+1}x. \]
The second two terms are $O(\bar{Y}^2)$, and since $\bar{Y}$ and its derivatives are small inside $D$ they can be neglected. The first two terms are evaluated by treating the background field as a small perturbation of the soliton field and retaining the constant and linear terms in $\bar{Y}$.  Inserting the expressions
\[ g=\tilde{g}\exp(\tilde{g}^{-1}\bar{Y}\tilde{g})\mbox{ and }\quad L_i \approx \tilde{L_i} + \pi_\m\left(\pa_i (\tilde{g}^{-1}\bar{Y}\tilde{g})+[\tilde{A}_i,\tilde{g}^{-1}\bar{Y}\tilde{g}]\right) \]
into the integral yields
\begin{align*}
I_2 &= I_3+I_4\\ 
I_3 &= -\int_{I\times D} \Big\{  \mathcal{V}(\tilde{L}_i) + \mathcal{U}(\tilde{g}) \Big\}\dd^{d+1}x \\
I_4 &= - \int_{I\times D} \Big\{\frac{\pa\mathcal{V}}{\pa L_i^a}(\tilde{L}_j)(\pa_i (\tilde{g}^{-1}\bar{Y}\tilde{g})^a + f_{\alpha b}^a \tilde{A}_i^\alpha(\tilde{g}^{-1}\bar{Y}\tilde{g})^b) + \delta_a\mathcal{U}(\tilde{g})(\tilde{g}^{-1}\bar{Y}\tilde{g})^a \Big\} \dd^{d+1}x
\end{align*}
Since $\tilde{g}$ is close to the vacuum outside $D$ the domain of integration of $I_3$ can to a good approximation be replaced by $I\times\RR^d$, so 
\begin{equation}\label{I3} I_3 = -\int_I M_S \dd x^0. \end{equation}
On integration by parts $I_4$ reduces to a boundary integral
\[ I_4 = - \int_{I\times \pa D} \Big\{\frac{\pa\mathcal{V}}{\pa L_i^a}(\tilde{L}_j)(\tilde{g}^{-1}\bar{Y}\tilde{g})^a\Big\}\dd\Sigma_i \]
because $\tilde{g}$ solves the Euler-Lagrange equation
\[ \frac{\pa}{\pa x^i}\frac{\pa\mathcal{V}}{\pa L_i^a}(\tilde{L}_j)- f_{\alpha a}^b\tilde{A}_i^\alpha\frac{\pa\mathcal{V}}{\pa L_i^b}(\tilde{L}_j) - \delta_a\mathcal{U}(\tilde{g}) = 0. \]

\noindent\textit{Step 3.}  The boundary integrals $I_1$ and $I_4$ will now be evaluated together.  From the approximate form \eqref{linearised lagrangian} of the lagrangian we obtain
\[ \frac{\pa \mathcal{L}}{\pa L_i^a}(\exp(Y)) \approx -\kappa_{ab}\pa_i Y^b,\quad \frac{\pa\mathcal{V}}{\pa L_i^a}(\exp(Y)) \approx \kappa_{ab}\pa_i Y^b \]
for small $Y$.  Substituting these into $I_1$ and $I_4$, approximating $\tilde{g}^{-1}\bar{Y}\tilde{g}$ by $\bar{Y}$ and using the fact that $\dd\Sigma_0=0$ on $I\times\pa D$ gives
\[ I_1+I_4 = \int_{I\times \pa D} \kappa_{ab} ( \tilde{Y}^a \pa_i\bar{Y}^b - \pa_i\tilde{Y}^a \bar{Y}^b) \dd \Sigma_i. \]
Replacing $\tilde{Y}$ with $\tilde{Q}^i\pa_i G$, where $\tilde{Q}^i=h_SQ^jh_S^{-1} R_j^i$, and applying the divergence theorem gives
\[ I_1+I_4 = \int_{I\times D}  \kappa_{ab}\tilde{Q}^{ia}\big(\triangle \bar{Y}^b\pa_iG - \bar{Y}^b\pa_i\triangle G \big) \dd^{d+1}x. \]
Substituting the Greens function equation $\triangle G(\mathbf{x}) = \delta(\mathbf{x})+m^2 G(\mathbf{x})$ and the equation of motion $\triangle \bar{Y} = \pa_0^2\bar{Y}+m^2\bar{Y}$ for $\bar{Y}$ gives
\[ I_1+I_4= \int_{I\times D}  \kappa_{ab}\tilde{Q}^{ia}\big( \pa_0^2\bar{Y}^b\pa_iG(\mathbf{x}) - \bar{Y}^b\pa_i\delta(\mathbf{x}) \big) \dd^{d+1}x. \]
The term involving $\pa_0^2\bar{Y}$ can be neglected since we have assumed that the derivatives of $\bar{Y}$ are approximately constant near the soliton.  Finally, integrating the remaining term by parts gives
\begin{align}
\nonumber
I_1+I_4
&= \int_{I\times D}  \kappa_{ab} \tilde{Q}^{ia} \pa_i\bar{Y}^b\delta(\mathbf{x})\,  \dd^{d+1}x \\
\label{I1I4}
&= \int_I \kappa_{ab}R_j^i(h_SQ^jh_S^{-1})^a\bar{L}_i^b \dd x^0.
\end{align}

\noindent\textit{Step 4.}  Now we evaluate the kinetic terms inside $D$, which are
\[ I_5 = \int_{D\times I} \big\{\mathcal{T}(L_0,L_i)-\mathcal{T}(\bar{L}_0,\bar{L}_i) \big\} \dd^{d+1}x. \]
We use the following expressions for $L_0$, $L_i$:
\begin{align*}
 L_0 &= \tilde{L}_0 + \pi_\m\big( \tilde{g}^{-1} \bar{L}_0 \tilde{g} \big) \\
 \tilde{L}_0 &= \pi_\m\big( \tilde{g}(\pa_0h_Sh_S^{-1}+\bar{A}_0)\tilde{g}^{-1}\big)-(\pa_0RR^{-1}\mathbf{x})^i\tilde{L}_i \\
 L_i &= \tilde{L}_i + \pi_\m(\tilde{g}^{-1} \bar{L}_i\tilde{g}) + \pi_\m(\tilde{g}^{-1}\bar{A}_i\tilde{g}^{-1}).
\end{align*}
Inserting these into $I_5$ gives
\begin{multline*}
I_5 = \int_{D\times I} \bigg\{\mathcal{T}(\tilde{L}_0,\tilde{L}_i) + \frac{\pa\mathcal{T}}{\pa L_0^a} (\tilde{L}_0,\tilde{L}_i) \big( \tilde{g}^{-1} \bar{L}_0 \tilde{g} \big)^a \\ + \frac{\pa\mathcal{T}}{\pa L_i^a} (\tilde{L}_0,\tilde{L}_i) \big( \tilde{g}^{-1}( \bar{L}_i+\bar{A}_i) \tilde{g} \big)^a \bigg\}\dd^{d+1}x
\end{multline*}
up to terms quadratic in the background fields.  Several of these terms vanish due to the integrands being odd functions. Since the second derivatives of $\bar{Y}$ are assumed to be small, we may treat $\bar{L}_\mu$ and $\bar{A}_\mu$ as constants over the domain of integration.  From this and the fact that $g_S(\mathbf{-x})=\sigma(g_S(\mathbf{x}))$ it follows that, as functions of $\mathbf{x}$:
\begin{itemize}
\item $\pi_\m\big( \tilde{g}^{-1} \bar{L}_0 \tilde{g} \big)$, $\tilde{L}_i$ and $\pi_\m(\tilde{g}^{-1} \bar{L}_i\tilde{g})$ are even; and
\item $\tilde{L}_0$ and $\pi_\m(\tilde{g}^{-1}\bar{A}_i\tilde{g}^{-1})$ are odd.
\end{itemize}
For purposes of illustration we will explain why $\tilde{L}_i$ is even; the others can be deduced by similar arguments.
The argument is:
\begin{align*}
\tilde{L}_i(-\mathbf{x}) &= -\pi_\m(\tilde{g}^{-1}(-\mathbf{x})\pa_i(\tilde{g}(-\mathbf{x})))\mbox{ by the chain rule} \\
&=-\pi_\m(\sigma\tilde{g}^{-1}(\mathbf{x})\pa_i(\sigma\tilde{g}(\mathbf{x})))\mbox{ by eq.\ \eqref{inversion symmetry}} \\
&=-\sigma\pi_\m(\tilde{g}^{-1}(\mathbf{x})\pa_i\tilde{g}(\mathbf{x})) \\
&= \pi_\m(\tilde{g}^{-1}(\mathbf{x})\pa_i\tilde{g}(\mathbf{x})) \\
&= \tilde{L}_i(\mathbf{x}).
\end{align*}
From the identity $\mathcal{T}(-\tilde{L}_0,\tilde{L_i})=\mathcal{T}(\tilde{L}_0,\tilde{L_i})$ and the above relations it follows that $\mathcal{T}(\tilde{L}_0,\tilde{L_i})$ is even, $\pa\mathcal{T}/\pa L_0^a(\tilde{L}_0,\tilde{L_i})$ is odd, and $\pa\mathcal{T}/\pa L_i^a(\tilde{L}_0,\tilde{L_i})$ is even.  It then follows that
\[
I_5 = \int_{D\times I} \bigg\{\mathcal{T}(\tilde{L}_0,\tilde{L}_i)  + \frac{\pa\mathcal{T}}{\pa L_i^a} (\tilde{L}_0,\tilde{L}_i) \big( \tilde{g}^{-1} \bar{L}_i \tilde{g} \big)^a \bigg\}\dd^{d+1}x.
\]
Both of these terms are quadratic in $\tilde{L}_0$.  The second is subleading because it is linear in the background field, whereas the first is independent of the background field.  Therefore we discard the second term.  After a change of variables $\mathbf{x}'= R^{-1}\mathbf{x}$ the first term gives
\begin{align}
\nonumber
I_5 &= \int_{D\times I} \mathcal{T}\Big(\pi_\m\big( g_S^{-1} h_S^{-1}(\pa_0h_S+\bar{A}_0h_S)g_S\big) - (R^{-1}\pa_0R\mathbf{x})^iL_i^S,\, L_i^S\Big)\dd^{d+1}x,
\end{align}
where $L_i^S=\pi_\m(g_S^{-1}\pa_i g_S)$.  To a good approximation the domain $D$ of integration can be replaced by $\RR^d$.  By comparing with equation \eqref{inertia tensor} we see that the leading contribution to $I_5$ is
\begin{equation}
\label{I5}
I_5 = \int_I \frac{1}{2}\Lambda_S\big(h_S^{-1}(\pa_0 h_S+\bar{A} h_S), R^{-1}\pa_0 R\big) \dd x^0.
\end{equation}

Finally, combining the results \eqref{I1I4}, \eqref{I3} and \eqref{I5} yields
\begin{align*}
S_B[\tilde{g}] &=I_1+I_3+I_4+I_5\\
&= \int_I\big\{ -M_S + \frac{1}{2}\Lambda_S\big(h_S^{-1}(\pa_0 h_S+\bar{A} h_S), R^{-1}\pa_0 R\big)\\
& \hspace{150pt} + \kappa_{ab}R_j^i(hQ^jh^{-1})^a\bar{L}_i^b \big\} \dd x^0 .
\end{align*}
This agrees with the general result \eqref{worldline action} in the case where $X^\mu(\tau)=(\tau,0,0,0)$.  Therefore \eqref{worldline action} holds for all paths with zero acceleration, by Poincar\'e invariance.

\subsection{Evaluation of the action: non-zero acceleration}

In order to evaluate the action on an accelerating soliton one must first write down a field $\tilde{g}$ that describes an accelerating path in the moduli space.  The simplest choice for $\tilde{g}$ is
\[ \tilde{g}(x^\mu) = h_S(t)g_S\big(R^{-1}(t)(\mathbf{x}-\mathbf{X}(t)\big) \]
as in \cite{schroers}.  We prefer not to use this path because it is incompatible with Lorentz invariance: for example, the field resulting from the constant velocity path $(h_S(t),R(t),\mathbf{X}(t))=(\mathrm{Id},\mathrm{Id},\mathbf{v}t)$ is not a Lorentz boost of a stationary soliton.  Inserting this field would result in an effective action which is not Lorentz-invariant, even though the field theory is Lorentz-invariant.  Instead, we propose to create a field $\tilde{g}$ using natural coordinates on the normal bundle of the soliton worldline.  This approach will be consistent with the simpler choice for $\tilde{g}$ when velocities are small, but has the added convenience of maintaining Lorentz invariance.

Assuming now that the worldline is parametrised by proper time $\tau$, the vector fields $\nu_i=m_i^\mu(\pa/\pa x^\mu)$ with $m_i^\mu$ defined in eq.\ \eqref{worldline boost} are an orthonormal frame for the normal bundle of the soliton worldline.  A natural set of coordinates $(\tau,\xi^i)\in\RR\times D$ in a tubular neighbourhood of the soliton worldine is given by
\[ x^\mu(\tau,\boldsymbol{\xi}) = X^\mu(\tau) + \xi^i m_i^\mu(\tau). \]
with $D\subset\RR^d$ a disc.  Assuming momentarily that this map is invertible, we define $\tilde{g}$ within the tubular neighbourhood by
\begin{equation}
\label{soliton in tubular neighbourhood}
\tilde{g}(x^\mu(\tau,\boldsymbol{\xi})) = h_S(\tau)g_S(R^{-1}(\tau)\boldsymbol{\xi}).
\end{equation}

With respect to these coordinates, the Minkowski metric takes the form
\[ -(1+\xi^i m_i^\mu\eta_{\mu\nu}a^\nu)^2\dd\tau^2 + (\dd\xi^i+\xi^j\Gamma^i_j\dd\tau)(\dd\xi^i+\xi^k\Gamma^i_k\dd\tau). \]
Here $a^\mu = \pa_\tau^2X^\mu$ is the acceleration and $\Gamma^i_j = \pa_\tau m_j^\lambda \eta_{\lambda\nu} m^\nu_\kappa\eta^{\kappa i}$.  Note that $\Gamma^i_j$ is proportional to the acceleration, so this metric looks like the Minkowski metric with acceleration-dependent corrections.  Assuming that these metric corrections are small, they can be neglected and the lagrangian density can be integrated over the tubular neighbourhood by repeating steps 2, 3, and 4 above.

The metric correction $\xi^i m_i^\mu\eta_{\mu\nu}a^\nu$ will be small only if the magnitude of the acceleration $a$ is considerably smaller than the reciprocal of the radius of the disc $D$.  Since in the previous calculation $D$ was required to be at least as large as the soliton, this means that the acceleration must be smaller than the reciprocal of the soliton size.    The condition that $a$ is small compared with the reciprocal of the disc radius also guarantees that the coordinates $(\tau,\boldsymbol{\xi})$ are well-defined functions of $x^\mu$.

Equation \eqref{soliton in tubular neighbourhood} only describes the soliton-carrying field $\tilde{g}$ inside the tubular neighbourhood; it remains to specify it outside the tubular neighbourhood.  We will not specify it explicitly, but simply assume that the field $\tilde{g}$ is close to the vacuum outside the tubular neighbourhood.  Under this assumption the integral of the lagrangian density outside the tubular neighbourhood can be reduced to an integral over the boundary of the tubular neighbourhood by treating the field $g=\bar{g}\tilde{g}$ as a small perturbation of the soliton field, as in step 1 above.  The result will depend only on the values of $\tilde{g}$ at the boundary of the tubular neighbourhood, which were given in \eqref{soliton in tubular neighbourhood}.  Thus it is not necessary for us to specify the precise form of $\tilde{g}$ away from the tubular neighbourhood.

To summarise, the action can be evaluated on an accelerating path in the soliton moduli space by inserting $\tilde{g}$ of eq.\ \eqref{soliton in tubular neighbourhood} into $S_B$.  The calculation will be very similar to that carried out above and the result will agree with \eqref{worldline action} provided that acceleration-dependent corrections are neglected.

\section{Quantisation}

In this section we treat the lagrangian \eqref{worldline action} quantum mechanically.  First we quantise the system canonically, obtaining a Schr\"odinger equation for a wavefunction on the moduli space.  Then, in the particular case of the Skyrme model, we apply perturbation theory and a Kaluza-Klein-type reduction to this Schr\"odinger equation.  The result of this process is a Schr\"odinger equation for a spinor-valued wavefunction.

\subsection{Canonical quantisation}

We begin our discussion of canonical quantisation by rewriting the lagrangian of eq.\ \eqref{worldline action} in more suitable form.  We write $\mathbf{X}$, $h_S$ and $R$ as functions of inertial time $t=x^0$ rather than proper time $\tau$.  We also rewrite the term involving the inertia tensor $\Lambda_S$.  Let $\kk\subset\h\oplus\mathfrak{so}(d)$ be the Lie algebra of $K\subset H\times\mathrm{SO}(d)$, and let $\kk^\perp\subset\mathfrak{h}\oplus\mathfrak{so}(d)$ be a $K$-invariant complement of $\kk$.  Let $E_A$ be a basis for $\kk^\perp$, where $A=1,\ldots,\dim(\kk^\perp)$.  Let $\Xi=(h_S,R)$ and let
\[ \Xi^{-1}D_t\Xi=(h_S^{-1}\pa_t h_S + h_S^{-1}(\bar{A}_0+\dot{X}^i\bar{A}_i)h_S,\,R^{-1}\pa_tR). \]
Since transformations in $K$ do not change the soliton, the kinetic energy due to rotation and isorotation is independent of the components of $\Xi^{-1}D_t\Xi$ in $\mathfrak{k}$.  Thus we may write
\[ \Lambda_S((h_S^{-1}D_th_S,R^{-1}\pa_tR) = \Lambda_{AB}(\Xi^{-1}D_t\Xi)^A(\Xi^{-1}D_t\Xi)^B, \]
where $(\Xi^{-1}D_t\Xi)^A E_A$ is the part of $\Xi^{-1}D_t\Xi$ in $\mathfrak{k}^\perp$ and $\Lambda_{AB}$ are components of the inertia tensor.

Using the approximations
\[ \dd\tau\approx (1-\sfrac{1}{2}|\dot{\mathbf{X}}|^2)\dd t,\quad \pa_\tau = (1+\sfrac{1}{2}|\dot{\mathbf{X}}|^2)\pa_t,\quad m_i^0=\pa_t X^i,\quad m_i^j=\delta_i^j \]
the lagrangian in \eqref{worldline action} is
\begin{multline}
\label{nonrelativistic lagrangian}
\mathcal{L}_B = -M + \sfrac{1}{2}M|\dot{\mathbf{X}}|^2 + \kappa_{ab}( h_SQ^ih_S^{-1})^a(\dot{X}^i\bar{L}_0^b + \bar{L}_i^b) + \\
 \sfrac12(1+\sfrac12|\dot{\mathbf{X}}|^2)\, \Lambda_{AB}(\Xi^{-1}D_t\Xi)^A(\Xi^{-1}D_t\Xi)^B .
\end{multline}

The textbook method of canonical quantisation involves a choice of local coordinates to act as position coordinates.  In our presentation $(\Xi,\mathbf{X})$ are not good coordinates on the moduli space, both because $\Xi$ is group valued rather than $\RR^n$-valued, and because of the equivalence $\Xi\sim\Xi k$ for $k\in K$.  Despite this, we will carry out the canonical quantisation treating $\Xi$ on a similar footing to a coordinate.  At the end of the procedure we will explain why quantising using local coordinates would give the same result (up to the usual operator ordering ambiguities).

The momenta dual to $\mathbf{X},\Xi$ are
\begin{align*}
P_i&:=\frac{\pa\mathcal{L}_B}{\pa \dot{X}^i} \\
&= \Big(M_S + \frac12 \Lambda_{\alpha\beta}(\Xi^{-1}D_t\Xi)^\alpha(\Xi^{-1}D_t\Xi)^\beta \Big)\dot{X}^i + \kappa_{ab} (h_SQ^ih_S^{-1})^a \bar{L}_0^b \\
&\quad + (1+\sfrac12|\dot{\mathbf{X}}|^2)\Lambda_{\alpha\beta}(h_S\bar{A}_ih_S^{-1})^\alpha (\Xi^{-1}D_t\Xi)^\beta \\
\Pi_A&:=\frac{\pa \mathcal{L}_B}{\pa (\Xi^{-1}\dot{\Xi})^A} \\
&=(1+\sfrac12|\dot{\mathbf{X}}|^2) \Lambda_{AB} (\Xi^{-1}D_t\Xi)^B.
\end{align*}
Here $A$ ranges from 1 to $\mathrm{dim}(\mathfrak{k}^\perp)$, so the number of momenta matches the dimension of position space.  We assume that the inertia tensor $\Lambda_{AB}$ is non-degenerate and write $\Lambda^{AB}$ for its inverse, such that $\Lambda^{AC}\Lambda_{CB}=\delta^A_B$.  Then the hamiltonian dual to $\mathcal{L}_B$ is
\begin{align*}
\mathcal{H} &:= P_i\dot{X}^i+\Pi_A (\Xi^{-1}\dot{\Xi})^A \\
&=M_S-\kappa_{ab} (h_SQ^ih_S^{-1})^a \bar{L}_i^b +\frac12\Lambda^{AB}\Pi_A\Pi_B - (h_S^{-1}\bar{A}_0h_S)^A\Pi_A \\
&\quad + \left(\frac{1}{2M}-\frac{\Lambda^{AB}\Pi_A\Pi_B}{4M^2}\right) \Big|\mathbf{P}-(h_S^{-1}\bar{\mathbf{A}}h_S)^A\Pi_A-\kappa_{ab} (h_S \mathbf{Q} h_S^{-1})^a\bar{L}_0^b \Big|^2
\end{align*}
up to terms of order four in either $P_i$ or $\Pi_A$ and of order 2 in $\bar{L}_\mu$.  Note that $(h_S^{-1}\bar{A}_ih_S)^A$ denotes the components of $h_S^{-1}\bar{A}_ih_S\in\h$ in $\mathfrak{k}^\perp\subset\h\oplus\mathfrak{so}(d)$.

Now we proceed to quantise this hamiltonian canonically.  The domain of the wavefunctions will be the configuration space $(H\times\mathrm{SO}(d))/K \times\RR^d$.  We will represent wavefunctions by functions $\Psi(\Xi,\mathbf{X})$ satisfying $\Psi(\Xi k,\mathbf{X})=\Psi(\Xi,\mathbf{X})$ for all $\Xi\in H\times\mathrm{SO}(d)$, $\mathbf{X}\in\RR^d$ and $k\in K$.  We make the standard substitutions
\begin{align}
\label{Phat}
\hat{P}_i\Psi(\Xi,\mathbf{X})&:= -\ii\hbar \frac{\pa\Psi}{\pa X^i},\\
\label{Pihat}
\hat{\Pi}_A\Psi(\Xi,\mathbf{X})&:= -\ii\hbar\frac{\dd}{\dd \epsilon} \Psi(\Xi\exp(+E_A\epsilon),\mathbf{X})\Big|_{\epsilon=0}.
\end{align}
Another operator that will appear in the quantised hamiltonian is
\begin{equation}
\label{Ahat}
\hat{A}_\mu\Psi:=  \frac{\dd}{\dd\epsilon} \Psi(\exp(-\epsilon\bar{A}_\mu)\Xi,\mathbf{X})\Big|_{\epsilon=0}.
\end{equation}
Note that $\hat{A}_\mu$ and $\hat{\Pi}_A$ commute, because the left- and right-actions of $H\times\mathrm{SO}(d)$ on itself commute.

With these substitutions made, the quantised hamiltonian takes the form
\begin{align}
\label{quantum hamiltonian}
\hat{\mathcal{H}} &= \hat{\mathcal{H}}_0 + \hat{\mathcal{H}}_1  \\
\label{free hamiltonian}
\hat{\mathcal{H}}_0 &:= \left(M_S+\sfrac{1}{2}\Lambda^{AB}\hat{\Pi}_A\hat{\Pi}_B\right) - \frac{\hbar^2|\nabla+\hat{\mathbf{A}}|^2}{2\left(M_S+\sfrac{1}{2}\Lambda^{AB}\hat{\Pi}_A\hat{\Pi}_B\right)} -i\hbar\hat{A}_0 \\
\label{interaction hamiltonian}
\hat{\mathcal{H}}_1 &:= -\kappa_{ab} (h_SQ^ih_S^{-1})^a \bar{L}_i^b +\frac{\ii\hbar \left\{ \kappa_{ab} (h_S Q^i h_S^{-1})^a\bar{L}_0^b,\,(\pa_i+\hat{A}_i) \right\}}{2\left(M_S+\sfrac{1}{2}\Lambda^{AB}\hat{\Pi}_A\hat{\Pi}_B\right)}
\end{align}
up to terms quadratic in $\bar{L}$ or quartic in $\hat{\Pi}_A$.  In deriving these expressions we have made use of the identity
\[
(h_S^{-1}\bar{A}_\mu h_S)^A\hat{\Pi}_A\Psi = \ii\hbar\hat{A}_\mu\Psi.
\]
This identity is proved as follows.  First note that
\[ 
\Xi\,\exp\big(\epsilon\pi_{\mathfrak{k}^\perp}(h_S^{-1}\bar{A}_\mu h_S)\big) = \Xi\,\exp\big(\epsilon h_S^{-1}\bar{A}_\mu h_S\big)\exp\big(-\epsilon\pi_\mathfrak{k}(h_S^{-1}\bar{A}_\mu h_S)+O(\epsilon^2)\big).
\]
Since $\Xi=(h_S,R)$ and $\bar{A}_\mu\in\h$,
\[ \Xi\,\exp\big(\epsilon h_S^{-1}\bar{A}_\mu h_S\big) = \big(h_S \exp\big(\epsilon h_S^{-1}\bar{A}_\mu h_S\big),R\big) = \exp\big(\epsilon\bar{A}_\mu\big)\,\Xi.\]
Therefore
\[ \Xi\,\exp\big(\epsilon\pi_{\mathfrak{k}^\perp}(h_S^{-1}\bar{A}_\mu h_S)\big) = \exp\big(\epsilon\bar{A}_\mu\big)\,\Xi\,\exp\big(O(\epsilon^2)\big)\,k \]
with $k=\exp(-\epsilon\pi_\mathfrak{k}(h_S^{-1}\bar{A}_\mu h_S))\in K$.  The identity follows from this and equations \eqref{Pihat} and \eqref{Ahat}.

This completes the calculation of the hamiltonian \eqref{quantum hamiltonian}.  As promised, we now explain why doing the calculation in local coordinates would lead to the same result.  Local coordinates on $(H\times\mathrm{SO}(d))/K$ are given by functions $Y^A(\Xi)$ on an open subset of $H\times\mathrm{SO}(d)$ satifying $Y^A(\Xi k)=Y^A(\Xi)$ for all $k\in K$.  The basic identity that we need is
\begin{equation}
\label{DAB}
\dot{Y}^A=D^A_B(\Xi^{-1}\dot\Xi)^B,\mbox{ where } D^A_B:=\frac{\dd}{\dd\epsilon} Y^A(\Xi\exp(\epsilon E_B))\Big|_{\epsilon=0}.
\end{equation}
This follows from
\[ \frac{\dd}{\dd t} Y^A(\Xi(t))\Big|_{t=t_0} = \frac{\dd}{\dd t} Y^A\Big(\Xi(t_0)\exp\big[(t-t_0)\Xi(t_0)^{-1}\dot{\Xi}(t_0)+O(t-t_0)^2\big]\Big)\Big|_{t=t_0}. \]
Let $S_A=\pa\mathcal{L}_B/\pa \dot{Y}^A$ be the momentum dual to $Y^A$.  It follows from eq.\ \eqref{DAB} that $\Pi_A=D_A^B S_B$ and hence that $\Pi_A(\Xi^{-1}\dot{\Xi})^A=S_B\dot{Y}^B$, so our classical hamiltonian $\mathcal{H}$ agrees with that defined using the coordinates $Y^A$.  To quantise the hamiltonian one should make the substitution $S_A\mapsto-\ii\hbar\pa/\pa Y^A$.  It follows from $\Pi_A=D_A^B S_B$ that $\Pi_A\mapsto D_A^B(-\ii\hbar\pa/\pa Y^B)$, and by the chain rule one can see that this operator agrees with $\hat{\Pi}_A$ defined in eq.\ \eqref{Pihat}.

\subsection{The case of the Skyrme model}
\label{subsec:skyrme}

In this section we describe how Schr\"odinger equation associated with the hamiltonian \eqref{quantum hamiltonian} reduces at low energies to a Schr\"odinger equation for a spinor-valued wavefunction on $\RR^3$ in the case of the Skyrme model.  Some parts of our description are simple generalisations of \cite{anw}.  We begin however with some general comments about the hamiltonian applicable in any field theory.

The hamiltonian \eqref{quantum hamiltonian} lends itself well to a perturbative treatment, with $\hat{\mathcal{H}}$ treated as a perturbation of $\hat{\mathcal{H}}_0$ by $\hat{\mathcal{H}}_1$.  Since $\Lambda^{AB}\hat{\Pi}_A\hat{\Pi}_B$ commutes with $\hat{\mathcal{H}}_0$, it is consistent to restrict $\hat{\mathcal{H}}_0$ to an eigenspace of $\Lambda^{AB}\hat{\Pi}_A\hat{\Pi}_B$.  The restriction of $\hat{\mathcal{H}}_0$ to the eigenspace corresponding to the least eigenvalue can be used to describe low-energy dynamics.  The restriction of $\hat{\mathcal{H}}_0$ in eq.\ \eqref{free hamiltonian} to this eigenspace is the hamiltonian for a charged particle, with the eigenvalue of $\Lambda^{AB}\hat{\Pi}_A\hat{\Pi}_B$ interpreted as a correction to the mass $M_S$ of the soliton.  One can then treat the perturbation $\mathcal{\hat{H}}$ of this ``free'' hamiltonian using standard methods.  The result will be an effective hamiltonian valid for low energies and weak background fields.

The accepted way to quantise the Skyrme model is through the Finkelstein-Rubinstein procedure \cite{fr68}.  In this procedure the wavefunction is not a function on the classical configuration space, but on its universal cover.  They impose the constraint that the values of the wavefunction at two distinct points in the universal cover corresponding to the same point in configuration space differ by a minus sign.  Finkelstein and Rubinstein showed that this procedure ensures that quantised solitons not only have half-integer spin but also enjoy Fermi exchange statistics.

Recall that the moduli space for the skyrmion is $(\mathrm{SU}(2)\times\mathrm{SO}(3))/\mathrm{SU}(2)\times \RR^3=\mathrm{SO}(3)\times \RR^3$.  The Finkelstein-Rubinstein procedure dictates that the wave function $\Psi$ is a function not on the moduli space but on its double cover $\mathrm{SU}(2)\times\RR^3$.  The covering map $\mathrm{SU}(2)\to (\mathrm{SU}(2)\times\mathrm{SO}(3))/\mathrm{SU}(2)$ is induced by
\[ h \mapsto \Xi = (h,\mathrm{Id}_3)\in\mathrm{SU}(2)\times\mathrm{SO}(3). \]
The Finkelstein-Rubinstein constraint on $\Psi:\mathrm{SU}(2)\times\RR\to \CC$ is that
\begin{equation}
\label{FR constraint}
\Psi(-h,\mathbf{X})=\Psi(h,\mathbf{X}).
\end{equation}

Now we evaluate some of the operators that appear in the free hamiltonian \eqref{free hamiltonian}.  The subalgebra $\mathfrak{k}\subset \mathfrak{su}(2)\oplus\mathfrak{so}(3)$ is spanned by $(\sigma_A/2\ii,J_A)$ and we choose $E_A=(\sigma_A/4\ii,\,-J_A/2)$ as a basis for its complement $\kk^\perp$.  It follows that the operators $\hat{\Pi}_A$, $\hat{A}_\mu$ appearing in \eqref{quantum hamiltonian} act on $\Psi(h,\mathbf{X})$ as
\begin{align*}
\hat{\Pi}_A\Psi(h,\mathbf{X}) &= -\ii\hbar\frac{\dd}{\dd\epsilon}\Psi\big(h\exp(\epsilon \sigma_A/2\ii),\mathbf{X}\big)\Big|_{\epsilon=0} \\
\hat{A}_\mu\Psi(h,\mathbf{X}) &= \frac{\dd}{\dd\epsilon}\Psi\big(\exp(-\epsilon \bar{a}_\mu)h,\mathbf{X}\big)\Big|_{\epsilon=0}.
\end{align*}
(the first of these follows from the identity
\[ (h,\mathrm{Id}_3)\exp(\epsilon E_A)=(h\exp(\epsilon\sigma_A/2\ii),\mathrm{Id}_3)k,\]
with $k=\exp(-\epsilon(\sigma_A/2\ii,J_A))\in K$).  From the formula \eqref{skyrmion inertia tensor} for the inertia tensor it follows that $\Lambda(Z^AE_A) = \Lambda_{AB}Z^AZ^B$, with
\[ \Lambda_{AB}=\frac{16\pi\hbar^2\lambda[f]}{3F_\pi e^3}\delta_{AB}. \]

The operator $\delta^{AB}\hat{\Pi}_A\hat{\Pi}_B$ is the Laplacian on $\mathrm{SU}(2)$ with its round metric.  Its spectrum is known from the Peter-Weyl theorem to be
\[ \{ \hbar^2\ell(\ell+1) \::\: \ell=0,1/2,1,3/2,\ldots\}. \]
The eigenspace $V_\ell$ with eigenvalue $\hbar^2\ell(\ell+1)$ is $(2\ell+1)^2$-dimensional; it has a basis given by the matrix entries of the irreducible $2\ell+1$-dimensional representation $\rho_\ell:\mathrm{SU}(2)\to \mathrm{U}(2\ell+1)$ (in other words, the spin-$\ell$ representation).  The Finkelstein-Rubinstein constraint \eqref{FR constraint} eliminates integer values of $\ell$ from the spectrum of $\hat{\Pi}_A\hat{\Pi}_A$, as was noted in \cite{anw}.  In particular, the smallest eigenvalue is $(1/2)\times(3/2)=3/4$, resulting in an effective mass in \eqref{free hamiltonian} of
\begin{equation}
\label{MN} M_N = M_S+\frac{1}{2}\Lambda^{AB}\hat{\Pi}_A\hat{\Pi}_B = \frac{\pi F_\pi \mu[f]}{e} + \frac{9F_\pi e^3}{128\pi\lambda[f]}.
\end{equation}
The corresponding wavefunction takes the form
\begin{equation}
\label{wavefunction} \Psi(\mathbf{x},h) = \Tr(\psi(\mathbf{x})h^{-1})
\end{equation}
for some $2\times 2$ matrix-valued function $\psi(\mathbf{x})$.  With this convention $\hat{A_\mu}\Psi=\Tr(\bar{a}_\mu\psi h^{-1})$, and $\hat{\mathcal{H}}_0\Psi(h,\mathbf{X})=\Tr(\hat{H}_0\psi(\mathbf{X}) h^{-1})$, where
\[ \hat{H}_0\psi = M_N\psi -\ii\hbar \bar{a}_0\psi - \frac{\hbar^2}{2M_N} |\nabla+\bar{\mathbf{a}}|^2\psi. \]

Now we calculate the first perturbative correction to this hamiltonian resulting from $\hat{\mathcal{H}}_1$.  The operator $\hat{\mathcal{H}}_1$ does not fix the subspace $V_{1/2}$, but if the background field is small it can be approximated by an operator $\pi_{1/2}\hat{\mathcal{H}}_1$ which does, where $\pi_{1/2}$ denotes orthogonal projection onto $V_{1/2}$.  The action of $\hat{\mathcal{H}}_1$ on $\Psi$ involves multiplication with the function
\[ (hQ^ih^{-1})^a = 4\pi q\left(\frac{2\hbar}{F_\pi g}\right)^2 R(h)^{ai} \]
of $h$, where $R(h)$ is the orthogonal matrix defined by $hI_bh^{-1}=R(h)^{ab}I_a$.  We claim that
\begin{equation}
\label{projection}
\pi_{1/2} (R^{ai}\Psi)(h,\mathbf{X}) = \frac{1}{3}\Tr(\sigma_a\psi(\mathbf{X}) \sigma_i h^{-1}).
\end{equation}

The identity \eqref{projection} is proved as follows.  First, note that $h\mapsto R(h)$ is the irreducible representation $\rho_1$ of $\mathrm{SU}(2)$.  There is a formula for $R(h)^{ai}$ (analogous to \eqref{wavefunction}):
\[ R(h)^{ai} = \Tr_{\CC^3}(E^{ai}\rho_1(h)^{-1}), \]
where $E^{ai}$ denotes the $3\times3$ matrix with 1 in the $a$th row and $i$th column, and zeros elsewhere.  Therefore
\begin{align*}
R(h)^{ai}\Psi(h,\mathbf{X}) &= \Tr_{\CC^2}(\psi(\mathbf{X})\rho_{1/2}(h^{-1}))\Tr_{\CC^3}(E^{ai}\rho_1(h^{-1})) \\
&= \Tr_{\CC^2\otimes\CC^3}\big( (\psi(\mathbf{X})\otimes E^{ai})(\rho_{1/2}(h^{-1})\otimes \rho_1(h^{-1}))\big).
\end{align*}
It is well-known that the tensor product $\rho_{1/2}\otimes\rho_1$ of representations is isomorphic to the direct sum $\rho_{1/2}\oplus\rho_{3/2}$ of irreducibles.  More precisely, there are unitary maps $p_{1/2}:\CC^2\otimes\CC^3\to \CC^2$, $p_{3/2}:\CC^2\otimes\CC^3\to\CC^4$ such that
\[ \rho_{1/2}(h^{-1})\otimes \rho_1(h^{-1}) = p_{1/2}^\dagger \rho_{1/2}(h^{-1}) p_{1/2} + p_{3/2}^\dagger \rho_{3/2}(h^{-1}) p_{3/2}\quad\forall h\in \mathrm{SU}(2). \]
Therefore
\begin{align*}
R(h)^{ai}\Psi(h,\mathbf{X}) &= \Tr_{\CC^2} \big(p_{1/2}(\psi(\mathbf{X})\otimes E^{ai}) p_{1/2}^\dagger\,\rho_{1/2}(h^{-1})\big) \\ 
& \qquad+ \Tr_{\CC^4} \big(p_{3/2}(\psi(\mathbf{X})\otimes E^{ai}) p_{3/2}^\dagger\,\rho_{3/2}(h_S^{-1})\big).
\end{align*}
The first summand on the right of this equation belongs to $V_{1/2}$ and the second to $V_{3/2}$.  Therefore the first term equals $\pi_{1/2}(R^{ai}\Psi)$.  To evaluate it one only needs to know the matrix entries for $p_{1/2}$; in terms of the Pauli matrices, and with respect to the standard bases for $\CC^2$ and $\CC^3$, these are $(p_{1/2})_{\alpha,(\beta i)}=(\sigma_i)_{\alpha\beta}/\sqrt{3}$ for $\alpha,\beta=1,2$ and $i=1,2,3$.  Note that the normalisation factor $1/\sqrt{3}$ is determined (up to an irrelevant phase) by the requirement that $(p_{1/2}p_{1/2}^\dagger)_{\alpha\beta}=\delta_{\alpha\beta}$.  It follows that $p_{1/2}(\psi\otimes E^{ai}) p_{1/2}^\dagger=\sigma_a\psi\sigma_i/3$, and the result \eqref{projection} follows.

It follows from \eqref{projection} that $\pi_{1/2}\hat{\mathcal{H}}_1\Psi(h,\mathbf{X})=\Tr(\hat{H}_1\psi(\mathbf{X})h^{-1})$, where
\[ \hat{H}_1\psi = \frac{4\pi\hbar q}{3g^2}\left(- \bar{L}_i^a\sigma_a\psi\sigma_i +  \frac{\ii\hbar}{2M_N} \big\{\bar{L}^a_0 \sigma_a ,\,\pa_i+\bar{a}_i\big\}\psi\sigma^i \right).  \]

For comparison with the chiral lagrangian it will be useful to convert the matrix $\psi(\mathbf{X})$ into a vector in $\CC^2\otimes\CC^2$.  To do so we write the matrix as $\psi^\alpha_\beta$, with $\alpha$ a row-index and $\beta$ a column-index, and let $\psi^{\alpha\beta}=\psi^\alpha_\gamma \epsilon^{\gamma\beta}$, where $\epsilon$ is antisymmetric and $\epsilon^{12}=1$.  One finds that $(-\psi\sigma_a)^{\alpha\beta}=\sigma_{a\gamma}^{\;\beta}\psi^{\alpha\gamma}$.  The vector $\psi^{\alpha\beta}$ can be regarded as a 2-component spinor transforming in the fundamental representation of the isospin group $\mathrm{SU}(2)$, with the first index $\alpha$ playing the role of an isospin index and the second index $\beta$ playing the role of a spin index.  In this way the Schr\"odinger equation $\ii\hbar\pa_t\psi = (\hat{H}_0+\hat{H}_1)\psi$ may be rewritten as
\begin{equation}
\label{skyrme schrodinger}
 \ii\hbar\frac{\pa\psi}{\pa t} = M_N\psi - \ii\hbar \bar{a}_0\psi - \frac{\hbar^2}{2M_N}|\nabla+\bar{\mathbf{a}}|^2\psi - \frac{4\pi\ii\hbar q}{3e^2}\bar{\ell}_i\sigma^i\psi - \frac{2\pi\hbar^2q}{3M_Ne^2}\big\{\bar{\ell}_0,\,\pa_i+\bar{a}_i\}\sigma^i\psi,
\end{equation}
with it being understood that $\bar{\ell}_\mu$ and $\bar{a}_\mu$ act on the isospin index of $\psi$ and the Pauli matrices act on the spin index.

\section{Comparison with the chiral effective lagrangian}

\subsection{The chiral effective lagrangian}

The chiral effective lagrangian is an effective lagrangian for pions and nucleons.  Pions are described by a field $U=\xi_L\xi_R^{-1}$ as in the Skyrme model, such that $U=\sigma\mathrm{Id}_2+\ii\pi^j\sigma_j$ with $\sigma^2=1-\pi^j\pi^j$, and nucleons are described by a four-component Dirac spinor $\Phi$ transforming in the fundamental representation of the isospin group $\mathrm{SU}(2)$.  The terms in the lagrangian are ordered according to the total number of derivatives and masses.  The first non-zero term involving nucleons is
\[ \mathcal{L}^{(1)}_{\pi N} = \bar{\Phi}(\ii\hbar\gamma^\mu(\pa_\mu+a_\mu)-M_N+\sfrac{\ii}{2}\hbar g_A\gamma^\mu\gamma_5\ell_\mu)\Phi. \]
We write the gamma-matrices as
\[ \gamma^0 = \begin{pmatrix} \mathrm{Id}_2 & 0 \\ 0 & -\mathrm{Id}_2 \end{pmatrix},\quad \gamma^i = \begin{pmatrix}0 & \sigma^i \\ -\sigma^i & 0 \end{pmatrix},\quad \gamma^5 = \ii\gamma^0\gamma^1\gamma^2\gamma^3 = \begin{pmatrix} 0 & \mathrm{Id}_2 \\ \mathrm{Id}_2 & 0 \end{pmatrix}, \]
and further write
\[ \Phi = \begin{pmatrix}\psi \\ \chi\end{pmatrix}. \]
Then the equations of motion for $\psi$ and $\chi$ are
\begin{align*}
0 &= (\ii\hbar( \pa_0+ a_0)-M_N+\sfrac{\ii}{2}\hbar g_A\sigma^j\ell_j)\psi + (\ii\hbar\sigma^j(\pa_j+a_j)+ \sfrac{\ii}{2}\hbar g_A \ell_0)\chi \\
0 &= (-\ii\hbar( \pa_0+ a_0)-M_N-\sfrac{\ii}{2}\hbar g_A\sigma^j\ell_j)\chi + (-\ii\hbar\sigma^j(\pa_j+a_j)-\sfrac{\ii}{2}\hbar g_A \ell_0)\psi.
\end{align*}
We suppose that $\ii\hbar(\pa_0+a_0)\Phi\approx M_N\Phi$ with $M_N\gg|\hbar g_A \ell_j|$ and solve the second equation approximately by writing
\[ \chi = -\frac{1}{2M_N} (\ii\hbar\sigma^j(\pa_j+a_j)+\sfrac{\ii}{2}\hbar g_A \ell_0)\psi. \]
Substituting this back into the first equation and neglecting terms quadratic in $\ell_\mu$ yields
\begin{equation}
\label{chiral schrodinger}
\ii\hbar\pa_0\psi = M_N\psi - \ii\hbar a_0\psi - \frac{\ii\hbar g_A}{2}\sigma^j\ell_j\psi - \frac{\hbar^2}{2M_N}(\pa_j+a_j)^2\psi - \frac{\hbar^2g_A}{4M_N}\{\ell_0,(\pa_j+a_j)\}\sigma^j\psi.
\end{equation}
Note that we have neglected a term proportional to $f_{ij}=\pa_i a_j-\pa_j a_i+[a_i,a_j]$; this is because the identity $\pa_{[\mu}(g^{-1}\pa_{\nu]}g)+(g^{-1}\pa_{[\mu}g)(g^{-1}\pa_{\nu]}g)=0$ implies that $f_{ij}$ is proportional to $[\ell_i,\ell_j]$ and hence quadratic in $\ell_\mu$.

The Schr\"odinger equations \eqref{skyrme schrodinger} and \eqref{chiral schrodinger} agree provided that
\begin{equation} \label{gA} g_A = \frac{8\pi q}{3e^2}. \end{equation}
This identification of parameters was also obtained in \cite{anw}.

\subsection{Calibration}
\label{subsec:calibration}

We now address the question of whether the correct values of the parameters $F_\pi,M_N,m_\pi,g_A$ that appear in the chiral effective lagrangian can be obtained by choosing the parameters $F_\pi,e,\bar{m}$ of the Skyrme model appropriately.  A particularly simple approach fix $F_\pi=185\mathrm{MeV}$ and tune the parameters $e$ and $m$ such that $g_A$ and $m_\pi$ 
equal the correct values of 1.29 and 137MeV.  One then obtains from \eqref{MN} a prediction for the nucleon mass.  Following this procedure, one obtains $e=3.57$ and $m=0.41$.  For this value of $m$ one has $\mu=12.1$, $\lambda=4.13$, $q=1.97$ and hence
\[ F_\pi=185\mathrm{MeV},\quad g_A=1.29,\quad m_\pi = 137\mathrm{MeV},\quad M_N=2016\mathrm{MeV}. \]
This value of the nucleon mass is clearly far too large.  One might seek to retune the Skyrme parameters so that $M_N$ is lower, but doing so will come at the cost of forcing the other parameters away from their experimental values.


Calibration problems of this type were encountered long ago \cite{anw}.  A possible resolution was found in \cite{moussallam93,hw95}.  These papers estimated the Casimir energy of the skyrmion, which is an $O(\hbar)$ correction to its mass.  Both found it to be negative and of magnitude around half of the classical skyrmion mass.  Thus including the Casimir contribution should result in a more acceptable value for the nucleon mass without altering the values of $g_A$, $m_\pi$ and $F_\pi$ -- indeed, this was conclusion of \cite{moussallam93,hw95}.  It would be profitable to revisit the calculation of the Casimir energy, as modern computing power might enable a more reliable estimate.

\section{Conclusion}

We have shown that the low energy dynamics of a quantised skyrmion are governed by the leading pion-nucleon term of the chiral effective lagrangian.  Consequently, chiral perturbation may be regarded as an effective description of skyrmion dynamics.

This result was derived under the assumptions that the skyrmion moves slowly and that the pion field with which it interacts is weak.  The latter condition requires that separations of individual skyrmions remain large compared with their radii.  Thus one might expect the Skyrme model and chiral perturbation theory to make similar predictions for long-range processes such as nuclear scattering.  On the other hand, the standard Skyrme model has classical bound states of skyrmions which differ radically from the well-separated skyrmions studied here, so the Skyrme model could provide insight into nuclear structure or dense nuclear matter which would be unattainable from chiral perturbation theory.

Our work suggestions several promising extensions.  Our quantum mechanical treatment of the Skyrme model in section \ref{subsec:skyrme} could be extended by including fields which take values in the space $V_{3/2}$ of eigenfunctions of the laplacian.  Doing so would result in a Schr\"odinger equation for coupled fields of spin 1/2 and 3/2 which could be compared with chiral lagrangians \cite{hwgs05} which couple the nucleon to the delta resonance.  It would also be interesting to try to recover subleading terms in the chiral effective lagrangian from the Skyrme model, but doing so would require more sophisticated methods than those presented here.  In particular, to be able to meaningfully compare parameters the Casimir effect may need to be included, as discussed in section \ref{subsec:calibration}.

From the point of view of the Skyrme model, it would be interesting to work out the implications of the Kaluza-Klein metric on the skyrmion moduli space for skyrmion scattering.  A lagrangian describing the dynamics of two well-separated skyrmions was calculated in \cite{schroers,gp94}.  This two-skyrmion lagrangian was shown in \cite{schroers} to agree with predictions of the lagrangian \eqref{worldline action} in the case where the gauge potential $\bar{A}_\mu$ vanishes.  Therefore the gauge potential in \eqref{worldline action} should lead to corrections to the lagrangian calculated in \cite{schroers,gp94}.  It would be interesting to work out the implications of these corrections for skyrmion scattering, particularly as recent numerical studies \cite{fk15} of skyrmion scattering have discovered behaviour at low impact parameter which deviates from the predictions of the lagrangian derived in \cite{schroers,gp94}.

\noindent\textbf{Acknowledgement}  We are grateful to the organisers of the 1st International Workshop on Nuclear Structure, Hadron Physics and Field Theory for providing a stimulating atmosphere, and to Nick Manton for raising the question addressed in this article.

\end{document}